\documentclass{sola}

\usepackage[hyphens]{url}
\usepackage{bm}
\usepackage{natbib}
\usepackage{graphicx}
\usepackage{amsmath}
\usepackage{multirow}

\begin{document}


%

\title{Zero-Shot Super-Resolution from Unstructured Data Using a Transformer-Based Neural Operator for Urban Micrometeorology}

%

\author{Yuki Yasuda\affil{1,2} and Ryo Onishi\affil{1,2}}

\affiliation{1}{Supercomputing Research Center, Institute of Integrated Research, Institute of Science Tokyo, Tokyo, Japan}
\affiliation{2}{d.weather inc., Tokyo, Japan}

\correspondingauthor{Yuki Yasuda}{Supercomputing Research Center, Institute of Integrated Research, Institute of Science Tokyo, Tokyo, Japan}{2-12-1 Ookayama, Meguro-ku, Tokyo 1528550, Japan}{yasuda.yuki@scrc.iir.isct.ac.jp}

%

\runningtitle{Zero-Shot SR from Unstructured Data Using a TNO}
\runningauthor{Yasuda and Onishi}


%

\begin{abstract}
This study demonstrates that a transformer-based neural operator (TNO) can perform zero-shot super-resolution of two-dimensional temperature fields near the ground in urban areas. During training, super-resolution is performed from a horizontal resolution of 100 m to 20 m, while during testing, it is performed from 100 m to a finer resolution of 5 m. This setting is referred to as zero-shot, since no data with the target 5 m resolution are included in the training dataset. The 20 m and 5 m resolution data were independently obtained by dynamically downscaling the 100 m data using a physics-based micrometeorology model that resolves buildings. Compared to a convolutional neural network, the TNO more accurately reproduces temperature distributions at 5 m resolution and reduces test errors by approximately 33\%. Furthermore, the TNO successfully performs zero-shot super-resolution even when trained with unstructured data, in which grid points are randomly arranged. These results suggest that the TNO recognizes building shapes independently of grid point locations and adaptively infers the temperature fields induced by buildings.
\end{abstract}

%

\section{Introduction} \label{sec:introduction}

Neural operators have recently been widely used in numerical simulations across the physical sciences \citep{Azizzadenesheli+2024, Liu+2025NeuroComp}. Neural operators approximate mappings from functions to functions (i.e., operators between function spaces) using neural networks, thereby enabling inference that is agnostic to the location of grid points. In physics, most quantities are represented as (continuous) ``fields,'' which are expressed as functions mapping spatial positions to physical quantities (e.g., $\bm{x} \mapsto T(\bm{x})$). Thus, neural operators are well suited for physics problems that describe the temporal evolution of fields via partial differential equations, and they have been applied to, for instance, numerical weather prediction \citep[e.g.,][]{Kurth+2023}.

Neural operators have also been applied to super-resolution \citep[SR; e.g.,][]{Wei+Zhang2023}. SR is a technique for enhancing data resolution and has been extensively studied using neural networks \citep{Lepcha+2023}. Neural operators regard images as continuous fields and formulate SR as a mapping from low-resolution fields to high-resolution fields. Since a continuous field is formulated as a function, such as $\bm{x} \mapsto T(\bm{x})$, neural operators for SR can perform inference at any grid point not included in training data. This approach allows neural operators to achieve arbitrary-scale SR \citep{Liu+2024}, which cannot be realized with conventional convolution-based networks that process pixel values on fixed, structured grids as arrays, not as continuous fields. As a result, neural operators can perform SR at higher resolutions than those used during training, an experimental setting known as zero-shot SR \citep[e.g.,][]{Wei+Zhang2023}.

SR using neural operators has recently started to be applied in meteorology \citep{Jiang+2023, Yang+2024, Sinha+2025}. These studies target meso- or global scales with horizontal resolutions of a few kilometers or coarser. In contrast, urban micrometeorology requires resolutions of a few meters to resolve building-induced flows \citep{Toparlar+2017}. Low-resolution simulations are computationally inexpensive but cannot resolve airflows, such as wind gusts, between buildings, whereas high-resolution simulations can resolve such small-scale motions but are computationally expensive. Previous studies have demonstrated the effectiveness of SR for inferring high-resolution flows from low-resolution data while reducing computational cost \citep{Onishi+2019, Wu+2021, Teufel+2023, Yasuda+Onishi2025}. However, these studies utilized conventional convolution-based networks, and the effectiveness of neural operator-based SR has not yet been established for urban micrometeorology.

This study employs a transformer-based neural operator (TNO) to demonstrate that zero-shot SR is possible for two-dimensional temperature fields near the ground in urban areas. The TNO can learn SR even from unstructured data with randomly arranged grid points and is considered capable of adaptively inferring the temperature fields induced by buildings.

\section{Methods} \label{sec:methods}

\begin{figure}
    \centering
    \includegraphics[width=14cm,angle=0]{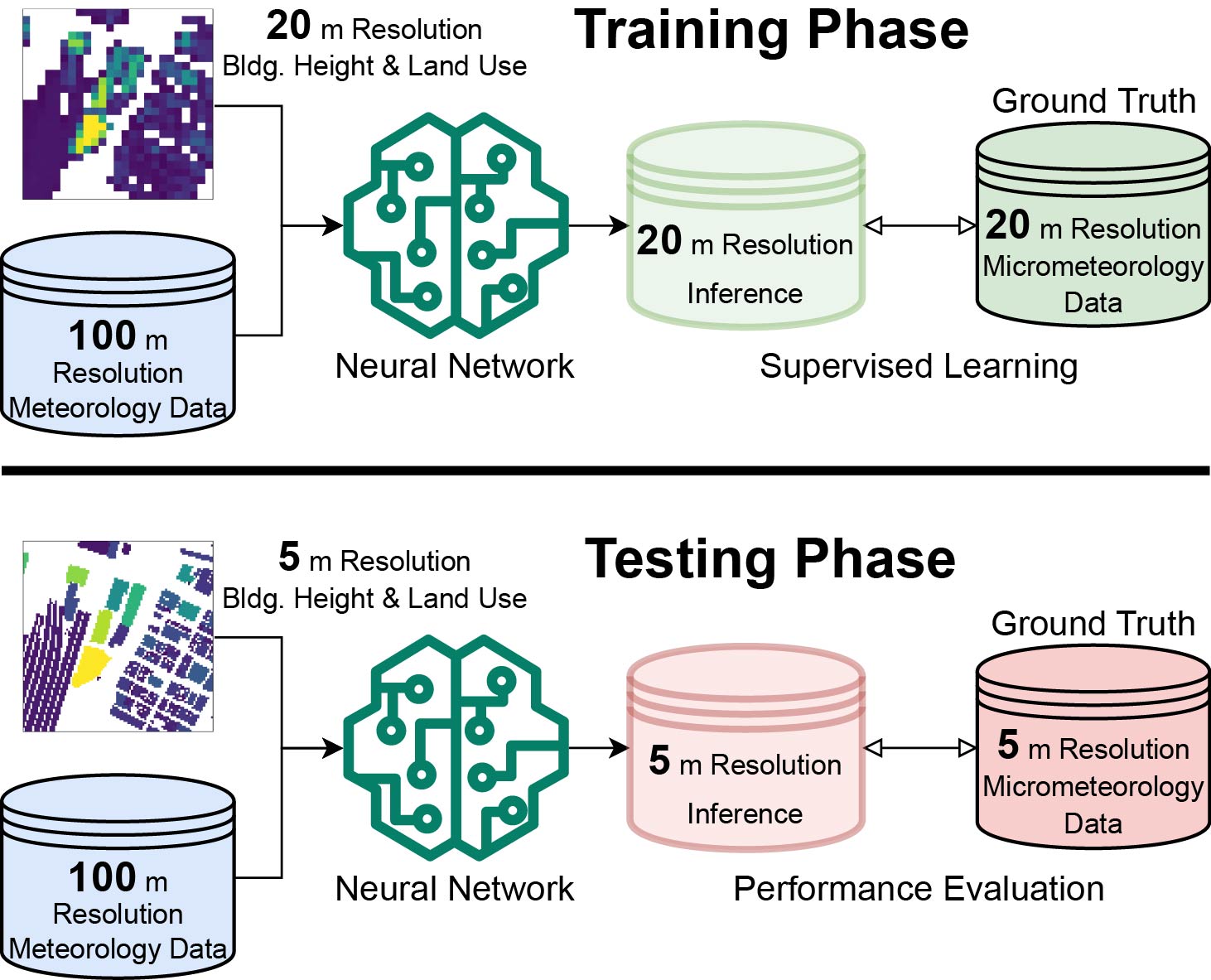}
    \caption{Experimental setup for zero-shot SR. Low-resolution 100 m data were dynamically downscaled using MSSG to obtain 20 m or 5 m data independently. In the training phase (upper panel), a neural network is trained to downscale from 100 m to 20 m resolution using ground truth at 20 m resolution. In the testing phase (lower panel), the trained network downscales from 100 m to 5 m resolution and is evaluated using ground truth at 5 m resolution. The 5 m resolution data are not used at all during the training phase.}
    \label{fig1}
\end{figure}

For the zero-shot SR in this study, inference was performed from horizontal resolutions of 100 m to 20 m during training, and from 100 m to 5 m during testing (Fig. \ref{fig1}), with the 5 m resolution data not used during training \citep[i.e., zero-shot configuration;][]{Azizzadenesheli+2024, Liu+2025NeuroComp}. These data are from simulations for extremely hot days from 2013 to 2020 \citep{Yasuda+Onishi2025}, obtained using a physics-based numerical model, the Multi-Scale Simulator for the Geoenvironment \citep[MSSG;][]{Onishi+2012, Takahashi+2013, Sasaki+2016, Matsuda+2018}. The 100 m resolution data were generated by mesoscale simulations, whereas the 20 m and 5 m data were independently obtained by dynamical downscaling of the 100 m resolution results (for data details, see \cite{Yasuda+Onishi2025}). The neural networks are trained to emulate this dynamical downscaling and then evaluated under the zero-shot setting. We show that this emulator significantly reduces the computation time of dynamical downscaling (Section 3).

\subsection{Data} \label{subsec:data}

\begin{figure}
    \centering
    \includegraphics[width=17cm,angle=0]{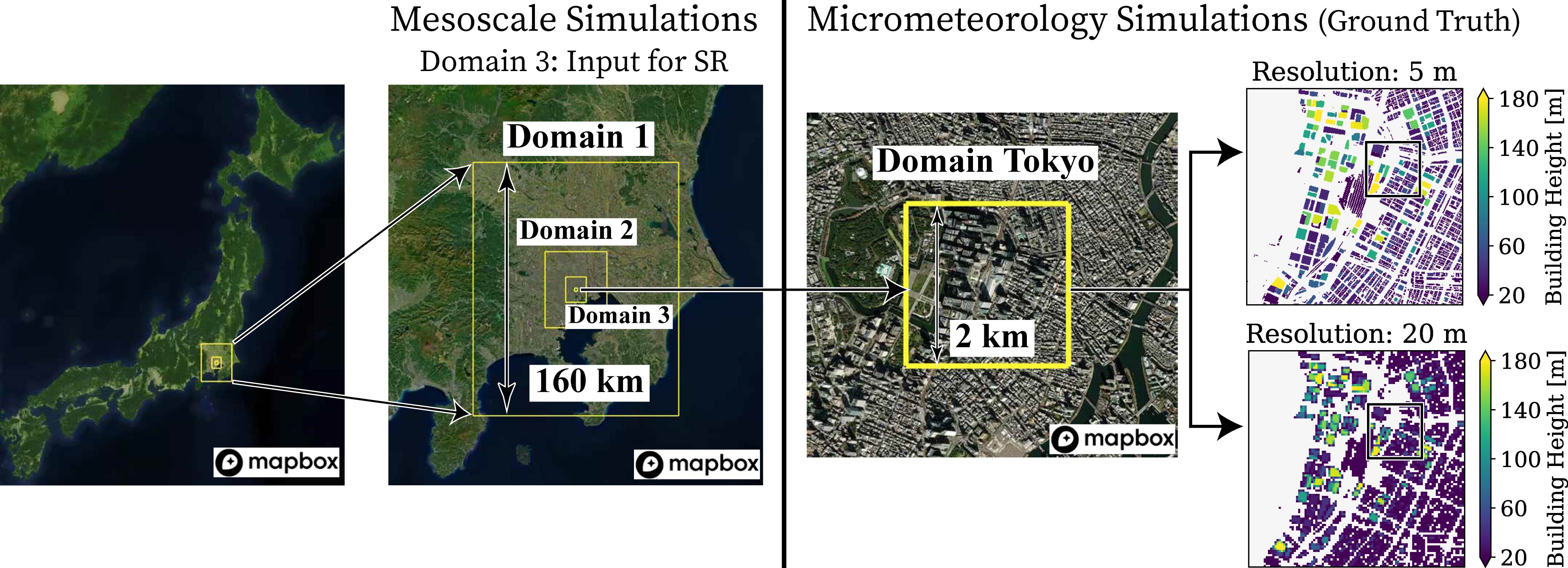}
    \caption{Computational domains for mesoscale and micrometeorology simulations. Dynamical downscaling from mesoscale to microscale was conducted using MSSG. The center of all domains is Tokyo Station in Japan. The black squares in the building height distributions indicate the regions shown in Figs. \ref{fig4} and \ref{fig5}. The satellite images were obtained from \cite{mapbox} and \cite{openstreetmap}. Figure adapted from Yasuda and Onishi (2025) under CC BY 4.0 license.}
    \label{fig2}
\end{figure}

Figure \ref{fig2} shows the computational domains of the MSSG simulations. The center of all domains is Tokyo Station in Japan. The mesoscale simulations used two-way nesting with three computational domains, where Domains 1, 2, and 3 have horizontal resolutions of 1 km, 300 m, and 100 m, respectively. The building-resolving micrometeorology simulations were conducted using one-way nesting, with the results of Domain 3 serving as boundary and initial conditions. Specifically, independent simulations were performed with uniform resolutions of 20 m or 5 m in all directions, and the resulting temperature at a height of 2 m was used as the ground truth.

The neural networks utilize the mesoscale simulation data as input. These consist of temperature and horizontal wind velocity at the lowest five levels, ground temperature, and vertically interpolated 2 m height temperature. Additionally, the input includes static data from the micrometeorology simulations: building height, land-use index, and the $(x,y)$ coordinates of output grid points. The ground truth (and the neural network output) is the 2 m height temperature at 20 m or 5 m resolutions. All variables cover a 1.6 km square area centered on Tokyo Station and are one-minute-averaged values from MSSG. The 100 m resolution inputs are upsampled by nearest-neighbor interpolation to 80 $\times$ 80 grids for the 20 m case or 320 $\times$ 320 grids for the 5 m case. The data from 2013–2019 were used for training (2,880 sets), and the data from 2020 were used for testing (540 sets). All results shown in Section \ref{sec:results-and-discussion} are based on the test data.

Unstructured grid data were also created to confirm that inference by neural operators is independent of the location of grid points \citep{Azizzadenesheli+2024, Liu+2025NeuroComp}. For each ground truth (20 m or 5 m resolution), we randomly scattered the same number of points as in the original gridded data and assigned values via bicubic interpolation. For the input at 100 m resolution, values at the same scattered points as the ground truth were also calculated using bicubic interpolation. These unstructured data can be directly fed to neural operators, but cannot be used as input to convolution-based networks that require regular grids.

\subsection{Neural networks} \label{subsec:neural-net}

\begin{figure}
    \centering
    \includegraphics[width=8cm,angle=0]{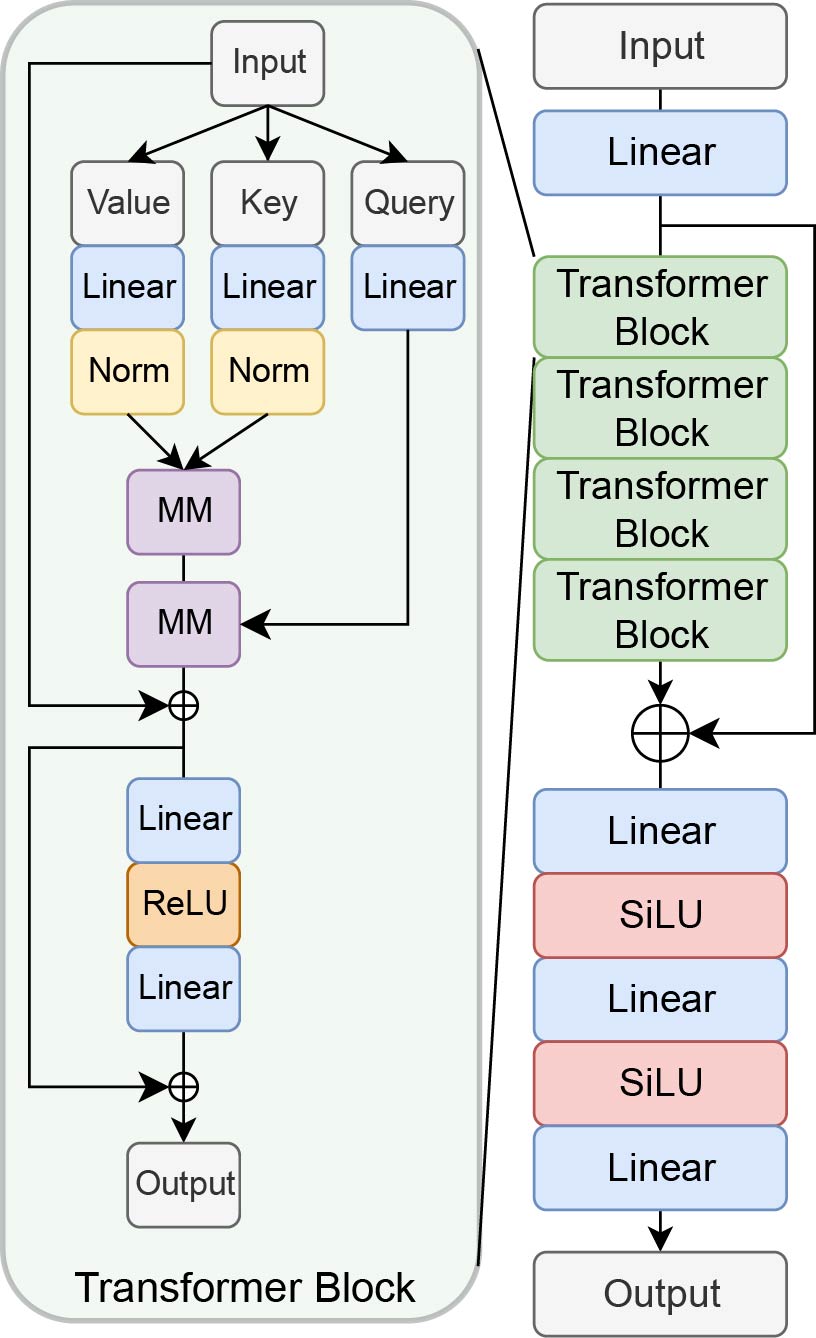}
    \caption{Transformer-based neural operator (TNO). ``Linear'' represents a linear transformation, ``SiLU'' represents the Sigmoid Linear Unit, ``Norm'' represents instance normalization \citep{Ulyanov+2016}, ``MM'' represents matrix multiplication, and ``ReLU'' represents the Rectified Linear Unit. The complete implementation is available in our Zenodo repository (see Data availability).}
    \label{fig3}
\end{figure}

Figure \ref{fig3} shows the transformer-based neural operator (TNO) used in this study. This model incorporates transformer blocks from \cite{Cao2021} and was developed based on an SR model for computer vision \citep{Wei+Zhang2023}. As a specific adaptation for urban micrometeorology, the building height and land-use index are added to a coordinate vector as additional channels at each output location. These additional data are static data at the target resolution (20 or 5 m). The complete implementation is available in our Zenodo repository (see Data availability).

The core transformer block \citep{Cao2021} can be viewed as an extension of the Fourier Neural Operator \citep[FNO;][]{Liu+2021}. The FNO was inspired by the pseudo-spectral method in computational fluid dynamics and achieves grid-independent inference by using Fourier transforms. The transformer block of \cite{Cao2021} is mathematically considered an extension of basis functions from trigonometric functions to learnable functions. Specifically, the matrix multiplication (MM) between ``Value'' and ``Key'' in Fig. \ref{fig3} corresponds to the Fourier transform, and the subsequent MM with ``Query'' corresponds to the inverse Fourier transform. \cite{Cao2021} proves the numerical stability of this transformer block; that is, the approximation error is bounded independently of the number of grid points.

As a comparison, we adopted a convolution-based network proposed by \cite{Yasuda+2022}. This model enhances SRCNN, an SR neural network \citep{Dong+2014}, with a channel attention mechanism \citep{Hu+2018}. The effectiveness of this model has been confirmed for inferring temperature fields in urban areas \citep{Yasuda+2022}. This convolutional neural network is simply referred to as CNN below.

The AdamW optimizer \citep{Loshchilov+Hutter2019} was used for training the TNO and CNN. The loss function for training was the mean absolute error (MAE; also known as L1 loss). MAE was selected instead of mean squared error (MSE; also known as L2 loss) to improve inference accuracy by avoiding splotchy artifacts in SR \citep{Zhao+2017IEEE}. For evaluation, we report two metrics: the MAE, which quantifies point-wise error, and the structural similarity index measure (SSIM) loss, which assesses pattern similarity error. For both metrics, smaller values indicate inferences closer to the ground truth. These evaluation metrics are widely adopted in SR research \citep{Lepcha+2023, Liu+2024}. The mean values of MAE and SSIM loss across all test data are referred to as M-MAE and M-SSIM loss, respectively. To evaluate the statistical significance of differences in M-MAE and M-SSIM loss, we conducted training five times with different initial weight parameters and calculated the standard deviations of both metrics.

We first compared inference performance by the CNN and TNO, both of which were trained on a mapping from structured (i.e., 100-m gridded) to structured (20-m gridded) data. These networks were then tested for mappings from structured (100-m gridded) to structured (20- or 5-m gridded) data. We next examined the TNO, which was trained on a mapping from unstructured (originally 100-m gridded) to unstructured (originally 20-m gridded) data. This TNO was tested for both unstructured-to-unstructured and structured-to-structured mappings. These experimental settings reveal a benefit of using unstructured data during training, namely, the enhancement of generalizability (Section \ref{sec:results-and-discussion}).

\section{Results and discussion} \label{sec:results-and-discussion}

\begin{figure}
    \centering
    \includegraphics[width=17cm,angle=0]{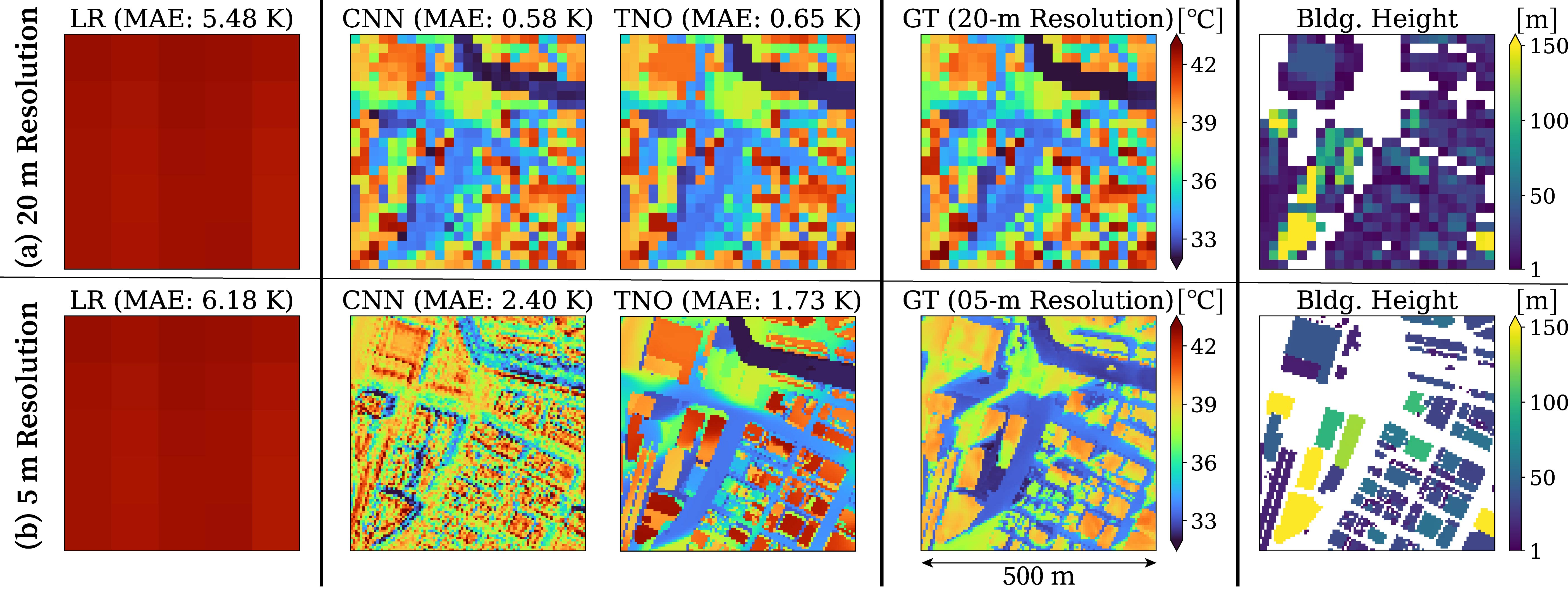}
    \caption{Example of SR results for temperature distributions at (a) 20 m and (b) 5 m resolution. The figures show a 500 m square area near Tokyo Station, indicated by the black squares in Fig. \ref{fig2}. ``LR'' represents the 2 m height temperature at 100 m resolution (low resolution), ``GT'' represents the 2 m height temperature ground truth, and ``Bldg. Height'' represents the building height distribution. ``LR'' refers to the same data in both rows, but the corresponding MAE differs because the GT has different resolutions of 20 m and 5 m.}
    \label{fig4}
\end{figure}

Figure \ref{fig4} shows an example of SR results. To validate successful training, the models were first evaluated using the 20 m resolution data from 2020. Both the CNN and TNO accurately reproduce the temperature field at 20 m resolution (Fig. \ref{fig4}a). This inference performs SR from 100 m to 20 m resolution (i.e., 5$\times$ SR), which is the same as during training. In contrast, for the zero-shot case (Fig. \ref{fig4}b), SR is performed to the 5 m resolution, which is finer than that used during training (i.e., 20$\times$ SR). Although the TNO exhibits a high-temperature bias, it qualitatively reproduces the pattern of the 5 m resolution temperature field well. The CNN generates noisy inference and fails to reproduce the temperature field induced by buildings. For instance, the CNN inference (Fig. \ref{fig4}b) shows noise over the L-shaped low-temperature region in the upper right (corresponding to an area above a river) and also fails to capture the low-temperature region caused by building shadows in the center. These results were consistently observed across other test data.

Table \ref{table1} shows the error values averaged over all test data. We first discuss the case where the training data are structured. For the test data at 20 m resolution, the CNN shows slightly smaller errors than the TNO. This result may be interpreted as the CNN having higher expressive power for fixed resolutions due to its larger number of parameters. Indeed, by increasing the number of transformer blocks in the TNO (i.e., increasing the number of parameters), we confirmed a decrease in error. This property---namely the improvement in expressive power by stacking transformer blocks---is also discussed in \cite{Cao2021}. Despite having about one-seventh as many parameters as the CNN, the TNO demonstrates higher generalizability. For the 5 m resolution data, the TNO shows smaller errors than the CNN, with the M-MAE and M-SSIM loss reduced by approximately 25\% and 31\%, respectively. The larger errors of the CNN reflect its noisy inference at 5 m resolution, as shown in Fig. \ref{fig4}b. These results indicate that conventional convolution-based models are sensitive to resolution and grid point locations, whereas neural operators are relatively insensitive to them, enabling zero-shot SR. These findings are consistent with previous results in computer vision \citep[e.g.,][]{Wei+Zhang2023, Liu+2024}.

\begin{figure}
    \centering
    \includegraphics[width=17cm,angle=0]{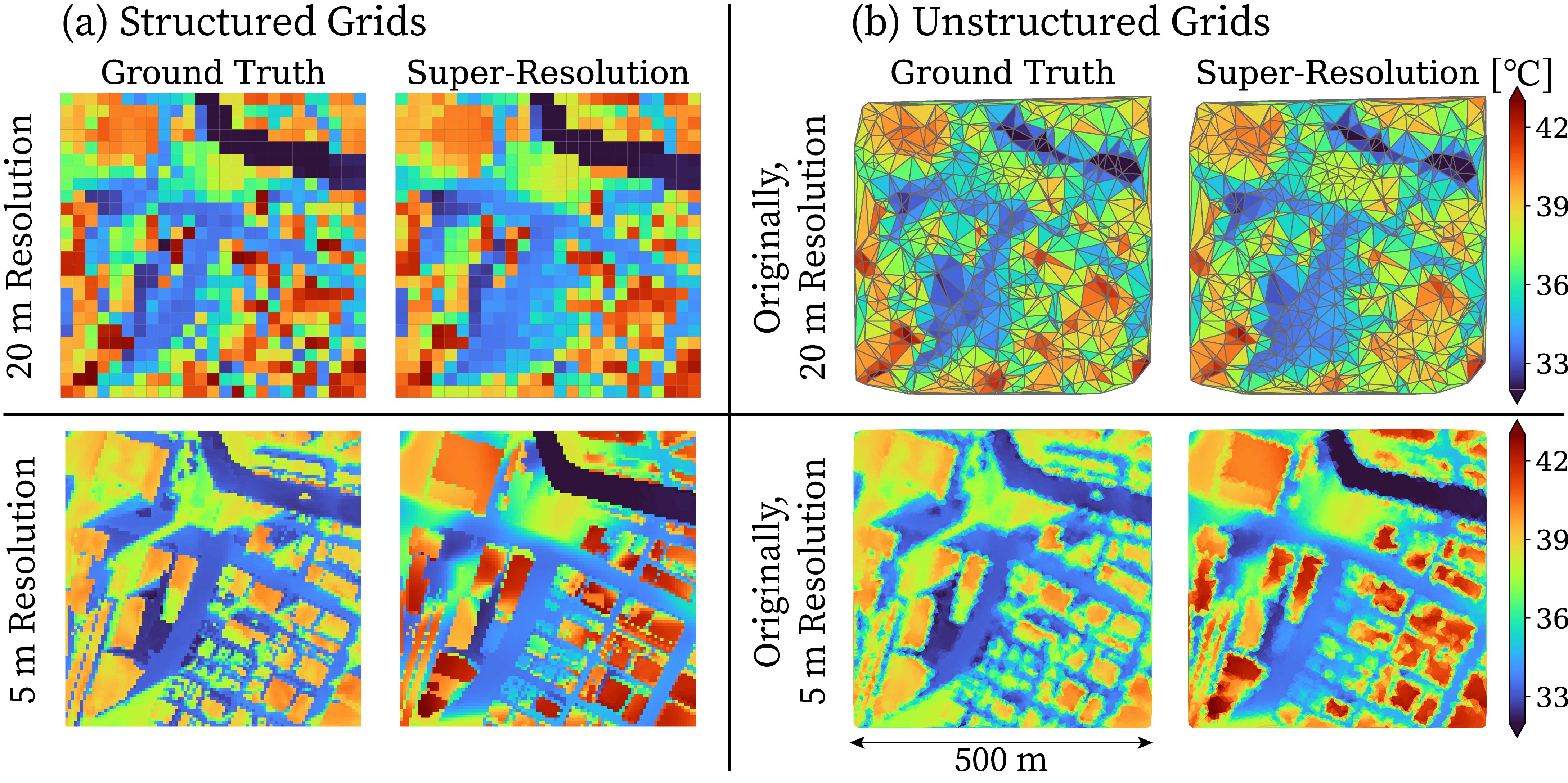}
    \caption{Example of test results from the TNO trained with unstructured data. In the upper panels, the ground-truth data were obtained from the 20 m resolution using (a) structured grids with 20 m spacing and (b) unstructured triangular grids. For (a), the same ground truth as in Fig. \ref{fig4} was used for this test. The grids are explicitly shown with gray lines. In the lower panels, the ground-truth data were obtained in the same manner using the 5 m resolution data instead of 20 m.}
    \label{fig5}
\end{figure}

Figure \ref{fig5} shows an example of inferences from the TNO trained using unstructured data. In Fig. \ref{fig5}a, inference is performed for equidistant grids (20 or 5 m spacing), which are identical to those in Fig. \ref{fig4}; whereas in Fig. \ref{fig5}b, inference is performed for the same type of unstructured grids as used during training. Note that we used the unstructured data obtained from the 20 m resolution temperature as the ground truth during this training. The TNO infers temperatures accurately at grid points specified for each test dataset, suggesting the independence of grid shapes. Indeed, the test errors for the equidistant grids do not strongly depend on the type of grids used during training (Table \ref{table1}). The M-MAE and M-SSIM loss are reduced by approximately 28\% and 38\%, respectively ($\sim$33\% on average), compared to those of the CNN. These error reduction rates are larger than those obtained using the structured grids (25\% and 31\%), implying that the use of unstructured grids during training enhances generalizability.

These results suggest that the TNO internally recognizes physical quantities as (continuous) ``fields'' \citep{Azizzadenesheli+2024, Liu+2025NeuroComp}. For ordinary images, neural operators may still require input of scale factors (i.e., ratios of low to high resolutions) \citep[e.g.,][]{Wei+Zhang2023}. In the present study, the TNO recognizes the shape of given buildings independently of grid point locations and adaptively infers the temperature field induced by these buildings (Fig. \ref{fig5}). Therefore, the TNO does not require input of the scale factor or structured input-output data. These findings may be specific to urban micrometeorology, where buildings characterize the flow scales.

Finally, we discuss the total inference time for 60-min predictions. We first report wall-clock times for physics-based simulations using MSSG, based on measurements taken on the Earth Simulator at the Japan Agency for Marine-Earth Science and Technology (JAMSTEC), which is equipped with AMD EPYC 7742 CPUs. On average, high-resolution (i.e., 5 m resolution) simulations took 206 min using 256 CPU cores, while low-resolution (i.e., 20 m resolution) simulations took 6.19 min using 40 CPU cores \citep{Yasuda+Onishi2025}. We measured the average training and inference times for SR on a local workstation equipped with an NVIDIA RTX 6000 GPU. Training the TNO required 613 min. Once trained, the TNO took 4.63 s for processing the 60-min low-resolution data (60 snapshots) regardless of whether the data were structured or unstructured. Consequently, the hybrid approach, which consists of the low-resolution simulation and the TNO inference, completed 60-min predictions of 2 m height temperature in 6.27 min, reducing the high-resolution computation time to 3.04\% (a 32.9 times speedup). This speedup factor is comparable to those reported in the latest surrogate modeling approaches for urban airflow simulations \citep{Shao+2023BAE, Peng+2024BAE}, although direct comparisons are difficult due to different experimental setups. In terms of computational cost (CPU-hours), the hybrid approach achieves a 213-fold reduction compared to the high-resolution simulation [$213 \approx (206 \times 256) / (6.19 \times 40)$].

\section{Conclusions} \label{sec:conclusions}

This study applied a transformer-based neural operator \citep[TNO;][]{Cao2021,Wei+Zhang2023} and demonstrated zero-shot super-resolution (SR) for two-dimensional temperature fields near the ground in urban areas. The test results showed that the TNO reproduces temperature distributions more accurately than a conventional convolutional neural network (CNN). The test errors from the TNO were approximately 33\% smaller than those from the CNN. The TNO also successfully learned SR even from unstructured data (i.e., randomly arranged point cloud data). These results suggest that the TNO adaptively infers the temperature fields induced by buildings.

Future research directions include improving accuracy and extending the method to three-dimensional data. For example, zero-shot learning \citep{Shocher+2018} could potentially improve accuracy while maintaining the zero-shot configuration (i.e., inference for unknown resolutions at test time). For practical applications, vertical structures of physical quantities are important, requiring SR of three-dimensional data. Since neural operators can be applied to three-dimensional data \citep[e.g.,][]{Qin+2025}, three-dimensional zero-shot SR is a promising next research direction in urban micrometeorology.

%

\section*{Acknowledgments}

The micrometeorology simulations were performed on the Earth Simulator system (project IDs: 1-23007 and 1-24009) at the Japan Agency for Marine-Earth Science and Technology (JAMSTEC). The deep learning experiments were conducted on a local workstation equipped with an NVIDIA RTX 6000 GPU board.

\section*{Data availability}

The source code for the deep learning models is preserved in the Zenodo repository (\url{https://doi.org/10.5281/zenodo.15307703}) and is openly developed in the GitHub repository (\url{https://github.com/YukiYasuda2718/zero-shot-sr-urban-mm}). The data that support the findings of this study are available from the corresponding author upon reasonable request.


%




\appendix
\section*{Appendix: Hyperparameters}

We list here the hyperparameters of the TNO, while the hyperparameters of the CNN are based on our previous study \citep{Yasuda+2022}. The main hyperparameters that we tuned are the learning rate and the number of transformer blocks. We confirmed that the results are not highly sensitive to these parameters. The other hyperparameters of the TNO are based on previous studies \citep{Cao2021, Wei+Zhang2023, Yasuda+2022}.

The number of features for all transformer blocks is 128, while it is increased to 256 in feed-forward networks after multi-head attention modules. The number of heads for these attentions is 8. The number of transformer blocks is 4. The AdamW optimizer \citep{Loshchilov+Hutter2019} was used with a learning rate of $2 \times 10^{-4}$ and a mini-batch size of 64. Each training was terminated using early stopping with a patience parameter of 50 epochs. The same stopping criterion was applied to the CNN.

%

\bibliographystyle{sola}
\bibliography{references}

%

%

\begin{table}
  \caption{Error values averaged over all test data (M-MAE and M-SSIM loss) for 20 m and 5 m resolutions, with columns indicating model type, grid type of training data, and number of parameters. All ground truth used for training was derived from the 2 m height temperature at 20 m resolution, regardless of whether the data were structured or unstructured. The values in parentheses are the standard deviations for M-MAE and M-SSIM loss. The abbreviation ``Res.'' stands for resolution.}
  \label{table1}
  \begin{center}
  \resizebox{\textwidth}{!}{
    \begin{tabular}{ccccccc}
      \hline
      \multirow{2}{*}{Model} & Grid Type of & Number of & M-MAE [K] & M-SSIM Loss & M-MAE [K] & M-SSIM Loss \\
      & Training Data & Parameters & (20 m Res.) & (20 m Res.) & (5 m Res.) & (5 m Res.) \\
      \hline
      CNN & Structured & 4,505,153 &  \textbf{0.57} ($\pm$ 0.01) & \textbf{0.029} ($\pm$ 0.001) & 2.21 ($\pm$ 0.05) & 0.485 ($\pm$ 0.015) \\
      TNO & Structured & 599,041 & 0.64 ($\pm$ 0.01) & 0.041 ($\pm$ 0.001) & 1.65 ($\pm$ 0.01) & 0.337 ($\pm$ 0.004) \\
      TNO & Unstructured & 599,041 & 0.75 ($\pm$ 0.02) & 0.055 ($\pm$ 0.003) & \textbf{1.60} ($\pm$ 0.02) & \textbf{0.302} ($\pm$ 0.001) \\
      \hline
    \end{tabular}
}
  \end{center}
\end{table}

\end{document}